\documentclass[aps,prl,twocolumn,notitlepage,superscriptaddress,showpacs,10pt,longbibliography,floatfix]{revtex4-2} 

\usepackage{amsmath}    
\usepackage{amsfonts}
\usepackage{bm}
\usepackage{amssymb}
\usepackage{graphicx}   
\usepackage[usenames,dvipsnames]{xcolor}      
\usepackage{comment}
\usepackage{siunitx}
\usepackage{enumitem}
\usepackage{upgreek}
\usepackage{color}
\usepackage{marvosym,amsthm}

\usepackage{soul}
\usepackage[normalem]{ulem}

\mathchardef\mhyphen="2D

\usepackage{xr}
\makeatletter

\newcommand*{\addFileDependency}[1]{
\typeout{(#1)}
\@addtofilelist{#1}
\IfFileExists{#1}{}{\typeout{No file #1.}}
}\makeatother

\newcommand*{\myexternaldocument}[1]{%
\externaldocument{#1}%
\addFileDependency{#1.tex}%
\addFileDependency{#1.aux}%
}

\myexternaldocument{supp}

\begin{document}

\title{Quantitative measure of topological protection in Floquet systems through the spectral localizer}

\author{Stephan Wong}
\email{stewong@sandia.gov}
\affiliation{Center for Integrated Nanotechnologies, Sandia National Laboratories, Albuquerque, New Mexico 87185, USA}\textbf{}

\author{Alexander Cerjan}
\email[]{awcerja@sandia.gov}
\affiliation{Center for Integrated Nanotechnologies, Sandia National Laboratories, Albuquerque, New Mexico 87185, USA}

\author{Justin T. Cole}
\email[]{jcole13@uccs.edu}
\affiliation{Department of Mathematics, University of Colorado, Colorado Springs, Colorado 80918, USA}

\date{\today}

\begin{abstract}
The standard understanding of topological protection from band theory is that a system's topology cannot change without first closing the bulk band gap. However, in Floquet systems, this typical definition of topological protection is one step removed from the experimentally accessible system parameters, as the relationship between the disorder in a system's instantaneous Hamiltonian and its Floquet Hamiltonian that defines its topology is not straightforward.
Here, we demonstrate that the spectral localizer framework for classifying material topology can be applied to Floquet systems and prove that its associated measure of topological protection can be understood in terms of the integrated disorder across the system's instantaneous Hamiltonians. As such, we have derived a quantitative bound on a Floquet system's topological protection in terms of the instantaneous system. Moreover, we show the utility of these bounds in both ordinary an anomalous Floquet Chern insulators.
\end{abstract}

\maketitle


In recent years, photonic Chern insulators have excited the photonics community due to their potential to yield robust wavelength-scale non-reciprocal devices, such as isolators and circulators.
While early work in realizing photonic Chern insulators utilized gyro-optical materials~\cite{Wang2009}, shifted ring resonators~\cite{Hafezi2011, Hafezi2013}, and exciton-polariton lattices~\cite{Klembt2018} to break, or effectively break, time-reversal symmetry, these platforms have yet to realize the full promise of wavelength-scale devices for nanophotonics applications, either because the magneto-optic effect is too weak at telecommunication wavelengths~\cite{Shintaku1994, Espinola2004, Bahari2017}, or because of the need for highly structured metasurfaces negates the flexibility provided by topological robustness against fabrication defects.
These challenges can instead be circumvented through the use of alternative approaches for breaking time-reversal symmetry; for example, it is possible to realize photonic Chern insulators by periodically driving an otherwise trivial system~\cite{DalLago2015, Leykam2016, Yan2015, Fang2019, Lu2021, Cheng2019}.
In particular, such topological Floquet systems have been proposed and experimentally demonstrated in helical waveguides array mimicking a time-periodic Hamiltonian~\cite{Rechtsman2013, Ablowitz2017, Maczewsky2017, Guglielmon2018, Ablowitz2019, Ivanov2019, Mukherjee2020, Maczewsky2020, Maczewsky2020a} or with periodically pump-driven lithium niobate photonic crystals~\cite{He2019}.

However, the characterization of topology in Floquet systems differs from that in static systems, as the system's time evolution must be incorporated into the classification framework.
Previously, the Floquet Hamiltonian, derived from the evolution operator, has been used to classify the topology of driven or otherwise time-periodic systems using tools from topological band theory~\cite{Cayssol2013}.
Nevertheless, although the Floquet Hamiltonian can be used to identify topologically non-trivial modes through its spectrum, topological band theory has been shown to be inadequate to correctly capture the topology in the topological anomalous Floquet modes, as the  associated Floquet bands have trivial Chern numbers~\cite{Rudner2013}.
To fully characterize the topology in Floquet systems, subsequent studies have argued that one needs to consider the full time-evolution of the unitary propagator~\cite{Nathan2015, Harper2020, Roy2017} or work with the time-reciprocal space Floquet Hamiltonian~\cite{Rudner2013}.
As such, these methods have yet to quantitatively resolve the robustness of topological modes in the presence of both spatial disorder and variations in the driving protocol.
Moreover, the use of a Floquet Hamiltonian to identify a system's topology obscures the effects of any topological protection. While topological band theory predicts that a system's topology can only change if sufficient perturbations are added to close the associated band gap, the Floquet bands of a driven system are abstracted from the instantaneous Hamiltonians that can be controlled in experiments. Thus, relating a quantitative definition of topological protection of Floquet systems to experimentally accessible parameters remains an outstanding challenge.

Here, we propose a framework based on the spectral localizer to diagnose the topology in Floquet systems directly from the Floquet Hamiltonian, and derive a bound for a system's topological protection in terms of perturbations to the system's associated time-dependent instantaneous Hamiltonian.
As the spectral localizer is able to capture the topology of any edge modes present in a spectrum, we demonstrate that the spectral localizer can also correctly identify the topology of the edge modes present in the Floquet bands, even if the system is an anomalous Floquet insulator.
Moreover, the local gap, which is a local and quantitative measure of the topological protection associated to the local topological markers obtained from the spectral localizer, provides insight into possible bounds on the spatial and temporal perturbation of the periodic driving protocol to preserve the non-trivial topology of the system.
In particular, we derive that the local topological markers cannot be changed as long as the time-average of the perturbations to the instantaneous system is smaller than the local gap determined using the Floquet Hamiltonian.
Overall, our classification framework lays out a general picture of the topology in Floquet system that can be directly related to the spectral localizer framework for static systems~\cite{Cerjan2022, Cerjan2022b, Cerjan2024, Dixon2023, Wong2023, Wong2024}, and provides insight into how careful the driving protocol should be in terms of perturbations.  
We anticipate that the Floquet spectral localizer framework and its associated quantitative measure of topological protection will be useful for experimental realizations of periodically driven photonic systems~\cite{Rechtsman2013, Maczewsky2017, Guglielmon2018, Maczewsky2020a}, as well as for driven nonlinear systems~\cite{He2019, Ivanov2019, Mukherjee2020, Maczewsky2020, Wong2023}. 
%



\begin{figure}
\centering
   \includegraphics[scale = 0.19]{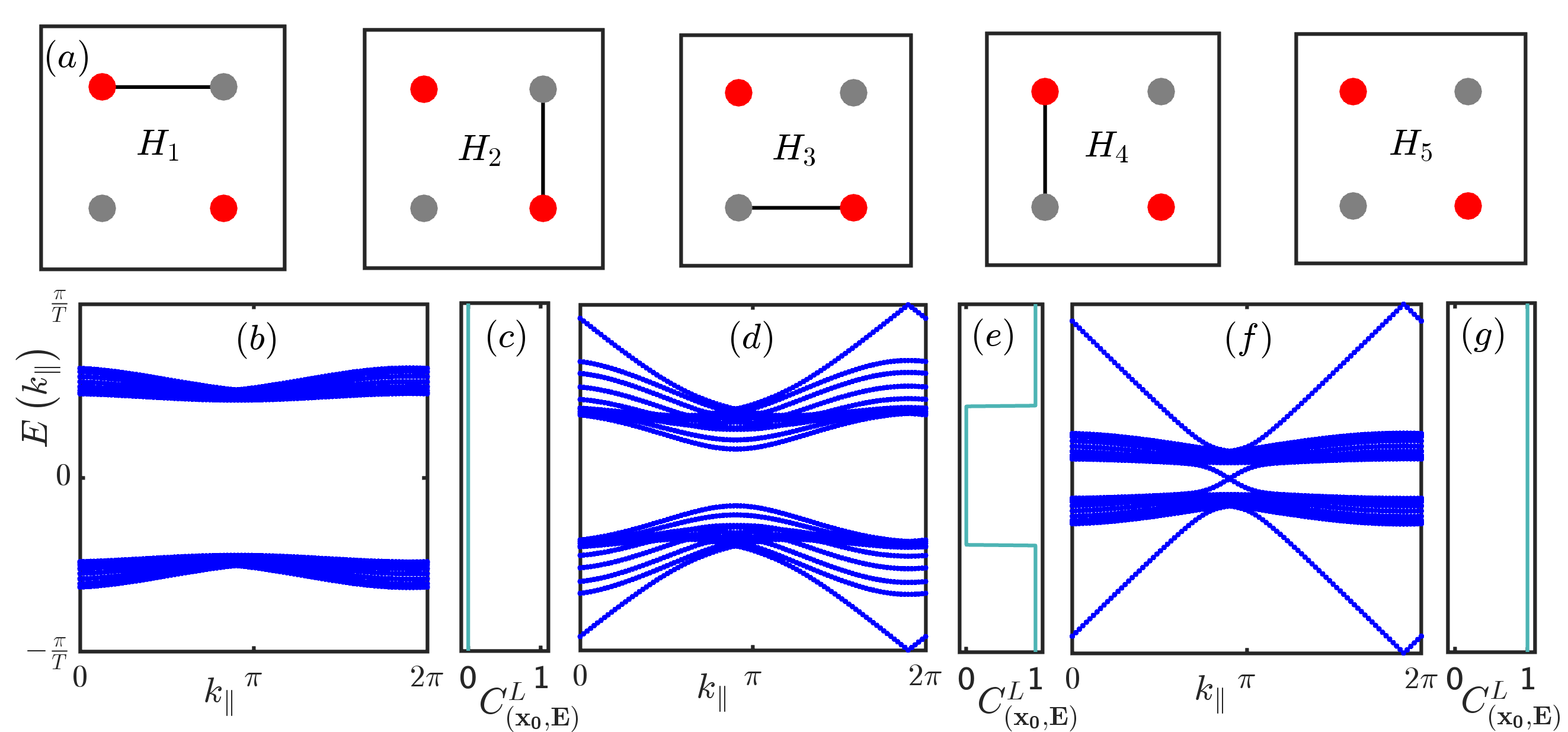}
    \caption{($a$) One period (5 subintervals) of the driving protocol. 
    Two sublattices are highlighted: grey dots ($a$-sites) and red dots ($b$-sites). Interactions of amplitude $J$ in the first four subintervals are indicated by lines; there is no hopping in the last subinterval. Over each subinterval a detuning term $\mp \Delta = \mp \frac{\pi}{2 T}$ is applied to the $a,b$-sites, respectively. (bottom row)  As the hopping amplitude increases, left-to-right, the system transitions from a trivial to anomalous Chern insulator. The bulk bands and the corresponding local index $C_{(\boldsymbol{x}_0,E)}^{\textrm{L}}$ given in Eq.~(\ref{local_Chern}) are shown for ($b,c$) $J = \frac{\pi}{2 T}$, ($d,e$) $J = \frac{3\pi}{2 T}$, and ($f,g$) $J = \frac{5\pi}{2 T}$. 
    The local Chern number has been calculating at the center $\boldsymbol{{x}_0}$ of a $10 \times 10$ lattice, with $\kappa = 0.5/(Ta)$, period $T=1 [a.u.]$ and lattice constant $a=1 [a.u.]$.
 \label{drive_protocol}}
\end{figure}

To illustrate how the spectral localizer framework can be applied to Floquet systems, we consider a standard system described by a 5-step driving protocol on a square lattice~\cite{Rudner2013}, as summarized in Fig.~\ref{drive_protocol}(a), with coupling strength $J$ and detuning mass term $\mp \Delta$ applied to the $a$ and $b$ sublattice sites, respectively.
As the coupling amplitude increases, the system goes through two topological transitions: from a trivial insulator to a Chern insulator, then from a Chern insulator to an anomalous Chern insulator, as shown by the existence or absence of edge states in the projected quasienergy bands in Fig.~\ref{drive_protocol}(b,d,f).
However, the bulk-edge correspondence from topological band theory is seemingly violated, because the bulk Chern number of the bands derived from the corresponding Floquet Bloch modes is unable to distinguish between the trivial and anomalous Chern insulators, with both band invariants being zero.

Instead, the topology of Floquet systems can be classified by generalizing the spectral localizer framework~\cite{Loring2015}.
The key advantage of this approach is that the spectral localizer 
takes an operator-based methodology to establishing topology, and as such is able to classify the local topology of any effective Hamiltonian and predict the existence (or absence) of topological protected modes regardless of the peculiarity of the system~\cite{Cerjan2022, Cerjan2022b, Dixon2023, Wong2024, Liu2023, Qi2024}.
In particular, consider a Floquet system 
%
\begin{equation}
\label{Floquet_system}
i \frac{d {\bf u}}{d t} = H(t) {\bf u}(t) , 
\end{equation}
%
characterized by a $N$-by-$N$ Hermitian Hamiltonian that is $T$-periodic $H(t + T) = H(t)$. 
The general solution of Eq.~\eqref{Floquet_system} is given by ${\bf u}(t) = \Pi(t,t_0) {\bf u}(t_0)$ where $\Pi(t,t_0)$ is the unitary evolution operator that can be expressed as $\Pi(t,t_0) = P(t,t_0)\exp(- i(t-t_0) H_\text{F} )$ such that $ P(t+T,t_0) = P(t,t_0)$ and $P(t_0,t_0) = I$ (identity)~\cite{Teschel}. 
In practice, given  linearly independent solutions $u_i(t), i=1,\ldots,N$ of Eq.~\eqref{Floquet_system}, the unitary evolution operator (matrix solution) can be written as 
%
\begin{equation}
\label{propagator_exp}
\Pi(t,t_0) = \left[ u_1(t) | \ldots | u_N(t) \right]
,
\end{equation}
%
and in the special case of $n$ piecewise constant subintervals, $\Pi(t,t_0)$ is simply the product of exponentials 
\begin{align}
\label{propagator_const}
\Pi(t_n,t_0) & = \Pi(t_n,t_{n-1})\cdots \Pi(t_2,t_1) \Pi(t_1,t_0)  \\  \nonumber
& = e^{-i (t_n - t_{n-1}) H_n } \cdots e^{-i (t_2 - t_1) H_2 } e^{-i (t_1 - t_0) H_1 } . 
\end{align}
%
For such Floquet systems, the effective Hamiltonian can be chosen to be the Floquet Hamiltonian $H_\text{F}$, obtained from the evolution operator after one period, i.e. the monodromy matrix $\Pi(t_0+T,t_0) = \exp( -iT H_\text{F})$, as~\cite{Loring2020b}
%
\begin{equation}
\label{Floquet_Hamil}
H_\text{F} = - \frac{\log  \Pi(t_0 + T,t_0)}{ i T}    
.
\end{equation}
%
The Floquet Hamiltonian is a Hermitian matrix (see Supplementary Material~\ref{extra_proof_append}~\cite{supp}), whose spectrum defines quasienergy bands that are similar to energy bands in static Hamiltonian, but unique only up to integer shifts of $2\pi/T$.
Compared to the instantaneous Hamiltonian $H(t)$, a system's Floquet Hamiltonian $H_\text{F}$ contains enough information of the periodic time-evolution in its quasienergy bands to enable the identification of a system's bulk and edge mode dispersion~\cite{Cayssol2013, Rudner2013}.

To diagnose the local topology of a two-dimensional Floquet system in class A~\cite{Altland1997, Schnyder2008, Kitaev2009, Ryu2010} at specific location $(x,y)$ and quasienergy $E$, the spectral localizer combines the Floquet Hamiltonian $H_\text{F}$ along with the position operators $X$ and $Y$ as 
%
\begin{align}
\label{define_floquet_localizer}
& L_{(x,y,E)}(X, Y, H_\text{F})  = \\ \nonumber
& \begin{pmatrix}
(H_\text{F} - E I) & \kappa (X - x I) - i \kappa (Y - y  I) \\
\kappa (X - x I) + i \kappa (Y - y  I) & - (H_\text{F} - E I) 
\end{pmatrix} ,
\end{align}
%
such that the system's local Chern number is given by
%
\begin{equation}
\label{local_Chern}
C_{(x,y,E)}^{\textrm{L}}(X, Y, H_\text{F}) = \frac{1}{2} {\rm sig} \left( L_{(x,y,E)}(X, Y, H_\text{F}) \right) \in \mathbb{Z} 
,
\end{equation}
%
where the signature ${\rm sig}$ of a matrix is its number of positive eigenvalues minus its number of negative eigenvalues.

In Eq.~(\ref{define_floquet_localizer}), $\kappa > 0$ is a hyperparameter used to make units consistent between the position and Hamiltonian operators as well as to balance the spectral weight between positions and energy, and is typically of the order of $\kappa \sim  E_\text{gap} / L$~\cite{Loring2020, Dixon2023, Cerjan2024} where $E_\text{gap}$ is the relevant spectral gap and $L$ the length of the finite system considered (see Supplementary Material~\ref{tuning_kappa_sec}~\cite{supp} for more discussion).
Moreover, depending on the system's dimension and the local symmetry of its Floquet Hamiltonian $H_\text{F}$ relative to the ten Altland-Zirnbauer classes~\cite{Altland1997, Schnyder2008, Kitaev2009, Ryu2010}, the spectral localizer can take different explicit formulations and the system's local topology at specific location and energy is given by either the signature, or the sign of the determinant or the Pfaffian of the spectral localizer~\cite{Loring2015, Kitaev2006}.
Overall, as the Floquet spectral localizer [Eq.~(\ref{define_floquet_localizer})]  can directly examine the topology in the system's bulk spectral gaps, our approach can distinguish between the trivial and the anomalous phases, even in a finite lattice. 
Indeed, Fig.~\ref{drive_protocol}(c,e,g) shows the energy-resolved local Chern numbers $C_{(x_0,y_0,E)}^{\textrm{L}}$ [Eq.~(\ref{local_Chern})] throughout the quasienergy band, demonstrating in accordance to the other topological Floquet theory~\cite{Rudner2013, Roy2017} that the anomalous Floquet phase is topological, despite having zero bulk Chern numbers.
%


The spectral localizer framework provides a rigorous treatment for predicting the system's topological protection.
As the local topological markers constructed from the spectral localizer are all tied to its spectrum~\cite{Loring2015}, the local gap, defined as 
%
\begin{equation}  
\label{local_band_gap}
\mu^\text{C}_{(\boldsymbol{x},E)}(\boldsymbol{X}, H_\text{F}) = \\ 
\min \big[ | {\rm spec} \left( L_{(\boldsymbol{x},E)}(\boldsymbol{X}, H_\text{F}) \right) | \big]
,
\end{equation}
%
gives a quantitative measure of the robustness of the topology; for a topological phase transition to take place, either as a result of system perturbations or changing the choice of $(\boldsymbol{x},E)$, the local gap must close, $\mu^\text{C}_{(\boldsymbol{x},E)}(\boldsymbol{X}, H_\text{F}) = 0$.
More precisely, consider the perturbed Floquet system
%
\begin{equation}
\label{perturbed_Floquet_eqn}
i \frac{d {\bf u}_{\rm pert}}{d t} = \left[ H(t) + \delta H(t) \right]{\bf u}_{\rm pert}(t) , 
\end{equation}
%
where $\delta H(t)$ is a possibly time-dependent perturbation on the instantaneous Hamiltonian in Eq.~\eqref{Floquet_system}, and is taken here to be a Hermitian and $T$-periodic matrix. 
Then the Floquet solution satisfies the evolution property ${\bf u}_{\rm pert}(T) = \exp(-i T H_\text{F,pert}) {\bf u}_{\rm pert}(0)$, where $H_\text{F,pert}$ is the (Hermitian) perturbed Floquet Hamiltonian.
Any change or perturbation in the Floquet Hamltonian $\delta H_\text{F} \equiv H_\text{F,pert} - H_\text{F}$ can be related to the local gap by
%
\begin{equation}
\left| \mu^\text{C}_{(\boldsymbol{x},E)}(\boldsymbol{X}, H_\text{F,pert}) - \mu^\text{C}_{(\boldsymbol{x},E)}(\boldsymbol{X}, H_\text{F}) \right| \le \Vert \delta H_\text{F} \Vert ,    \label{eq:10}
\end{equation}
%
through an application of Weyl's spectral theorem~\cite{Bhatia1997, Wong2023}, where $\Vert \cdot \Vert$ is the largest singular value of the matrix. 
As such, a topological transition cannot occur as long as 
%
\begin{equation}
\Vert \delta H_\text{F} \Vert <  \mu^\text{C}_{(\boldsymbol{x},E)}(\boldsymbol{X}, H_\text{F})
,     
\end{equation}
%
as the perturbation is not strong enough to close the system's local gap so that the local topological marker can change its value.

While Eq.~(\ref{eq:10}) yields a quantitative bound for the system's topological protection, this bound requires knowledge of the Floquet Hamiltonian, which may be challenging to determine in experimentally realizable systems.
Instead, a more convenient formulation would work directly with the instantaneous perturbations $\delta H(t)$ in Eq.~\eqref{perturbed_Floquet_eqn}, since these instantaneous perturbations are a measurable quantity that can be readily modeled or bounded in experiments and realistic simulations.
For sufficiently small perturbations, namely $\int_0^T  \Vert \delta H(t)  \Vert dt < 1$, the difference in the Floquet Hamiltonians is bound by the time-averaged perturbation Hamiltonian (see Supplementary Material~\ref{bound_dQ_sec}~\cite{supp})
%
\begin{equation}
\label{bound_instant}
\Vert \delta H_\text{F} \Vert \le   \overline{\sigma}_1  + \mathcal{O}\left(  \overline{\sigma}_1^2 T   \right)    + \mathcal{O}\left(  \frac{1}{T}  \right) 
, 
~
\overline{\sigma}_1  = \frac{1}{T} \int_0^T \Vert \delta H(t) \Vert dt 
.
\end{equation}
%
In the commutative case where $[H_\text{F,pert},H_\text{F}] = 0$, the $\mathcal{O}(T^{-1})$ term in Eq.~(\ref{bound_instant}) is identically zero.
Thus, for a Floquet system in a given Chern phase at $E$ indicated by $C_{(\boldsymbol{x},E)}^{\textrm{L}}(\boldsymbol{X},H_\text{F})$,  
%
\begin{multline}
\overline{\sigma}_1 < \mu^\text{C}_{(\boldsymbol{x},E)}(\boldsymbol{X},H_\text{F}) 
\\ 
\label{top_protect_cond}
\implies
C_{(\boldsymbol{x},E)}^{\textrm{L}}(\boldsymbol{X},H_\text{F,pert}) = C_{(\boldsymbol{x},E)}^{\textrm{L}}(\boldsymbol{X},H_\text{F})
.
\end{multline}
%
In other words, if the average disorder over one period is sufficiently small, then the topological phase at $C_{(\boldsymbol{x},E)}^{\textrm{L}}(\boldsymbol{X},H_\text{F})$ is preserved in the presence of $\delta H(t)$. 
Notice that Eq.~\eqref{top_protect_cond} represents a sufficient condition for guaranteeing the persistence of a topological (or trivial) phase, however in practice we find it is often more stringent than necessary for uncorrelated perturbations to preserve that phase. 
%

\begin{figure}
\centering
   \includegraphics[scale = 0.26]{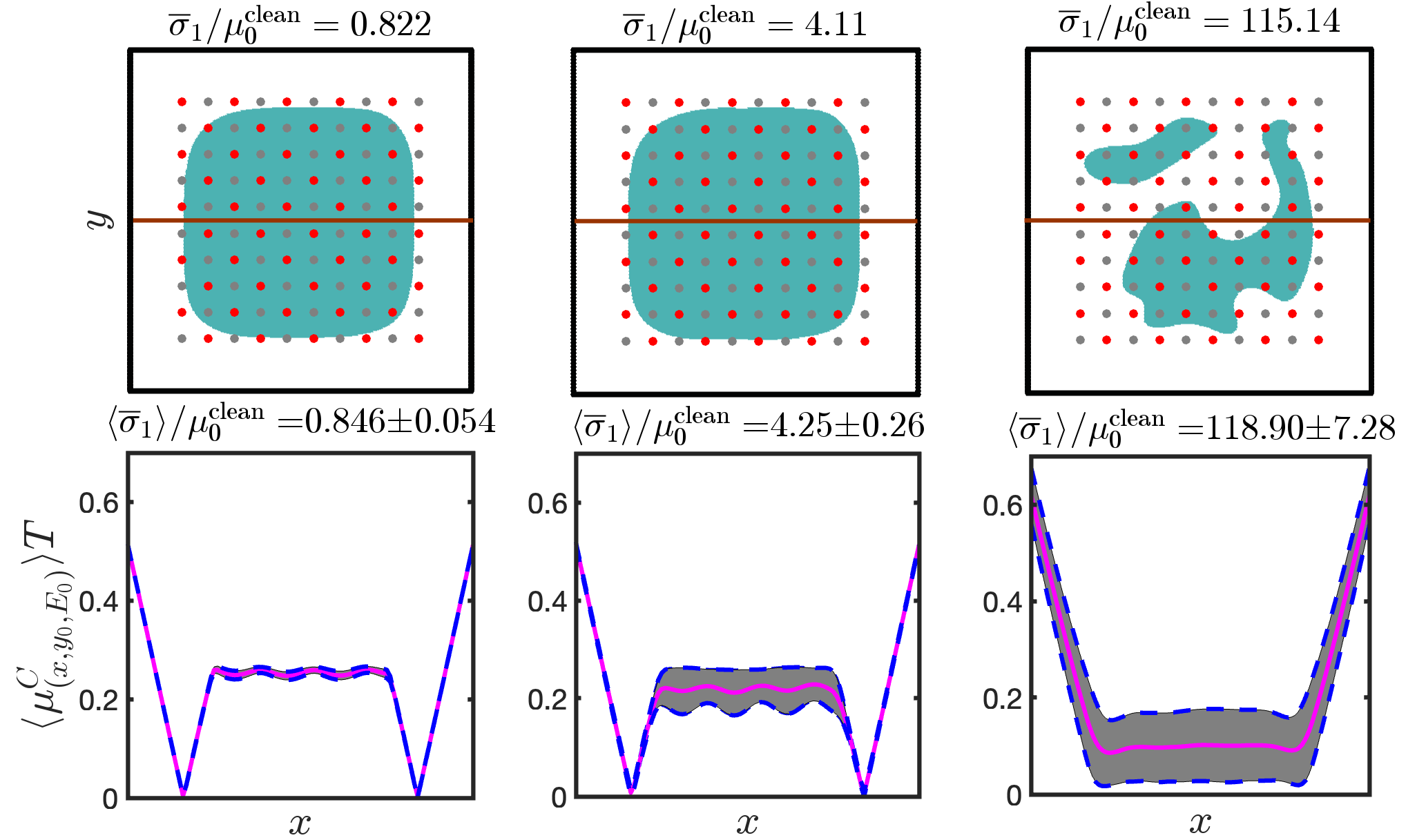}
    \caption{(top row) One realization of the local $0$-gap  Chern number $C^L_{(\boldsymbol{x},E_0)}$  ($E_0 = 0$)  (\ref{local_Chern})  as the strength of on-site disorder is increased, left-to-right,  in the anomalous insulator. The brown line indicates the  location  of $y_0$. (bottom row) Ensemble averaged local gap width (\ref{local_band_gap}) at $y_0$, with $\kappa = 0.5 /(Ta)$. Error bars denoting one standard deviation are included. 
 \label{zero_gap_onsite_disorder}}
\end{figure}

\begin{figure}
\centering
   \includegraphics[scale = 0.26]{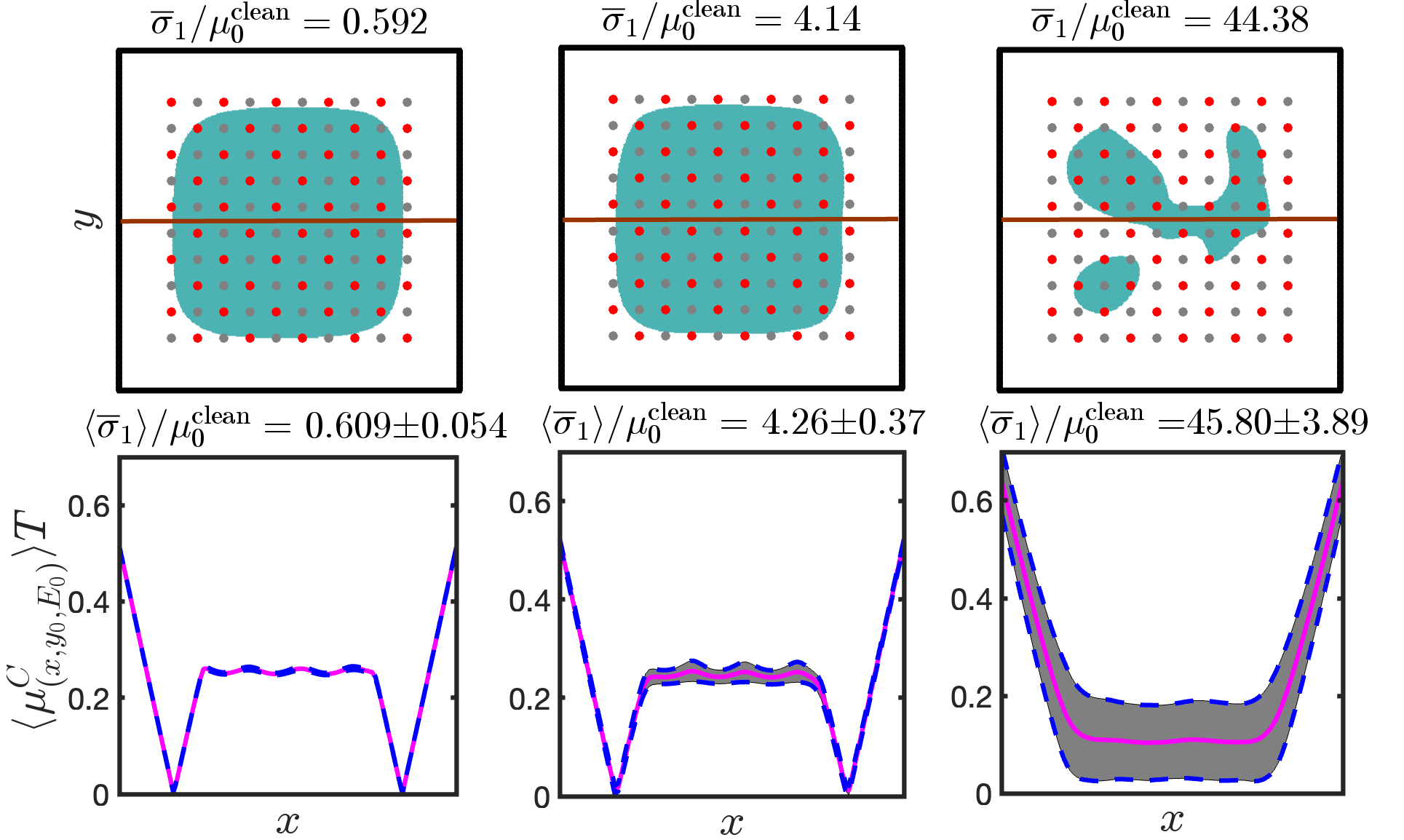}
    \caption{(top row) One realization of the local $0$-gap  Chern number  $C^L_{(\boldsymbol{x},E_0)}$ ($E_0 = 0$) (\ref{local_Chern})  as the strength of coupling disorder is increased, left-to-right,  in the anomalous insulator.  The brown line indicates the  location  of $y_0$. (bottom row) Ensemble averaged local gap width (\ref{local_band_gap}) at $y_0, \kappa = 0.5 / (Ta)$. Error bars denoting one standard deviation are included.
 \label{zero_gap_couple_disorder}}
\end{figure}

The derived bound, Eq.~\eqref{top_protect_cond}, serves as a guide for identifying the degree of robustness at various space-energy locations, relating topological protection of Floquet systems to experimentally accessible parameters. 
To explore the topological protection in a toy model system from Fig.~\ref{drive_protocol}, we first consider uncorrelated disorders, namely we introduce on-site and coupling disorders drawn from a normal distribution with mean zero and standard deviation $\gamma \Delta$ and $\gamma J$, respectively, while keeping $\delta H(t)$ Hermitian (see Supplementary Material~\ref{perturb_Hamil_sec}~\cite{supp}).
In waveguide systems, the former represents a perturbation of self-interaction $\Delta$, i.e., a variation in the local refractive index. 
The latter type of disorder can be used to model perturbations to the coupling coefficient $J$, e.g. inconsistent coupling strengths between pairs of waveguides. 
As the strength of the disorder is increased, quantified by $\overline{\sigma}_1$ [Eq.~(\ref{bound_instant})], the topologically nontrivial region degrades along the boundaries and ultimately ruptures into smaller domains, as shown in  Figs.~\ref{zero_gap_onsite_disorder}~and~\ref{zero_gap_couple_disorder}. 
When the disorder strength is smaller than the local gap of the unperturbed system $\mu^{\text{clean}}_{0} \equiv \mu^{\textrm{C}}_{(\boldsymbol{x}_0,E_{0})}(\boldsymbol{X},H_\text{F})$, namely $\overline{\sigma}_1 / \mu^{\text{clean}}_{0} < 1$, the topology of the system remains unchanged, thereby demonstrating the validity of derived bound in Eq.~\eqref{top_protect_cond}.
Nevertheless, the persistence of the non-trivial topology at larger disorder strengths, $\overline{\sigma}_1 / \mu^{\text{clean}}_{0} > 1$, illustrates that the inequality in Eq.~\eqref{top_protect_cond} serves as a sufficient, but not necessary, condition for preserving the topology.
%

\begin{figure}
\centering
   \includegraphics[scale = 0.325]{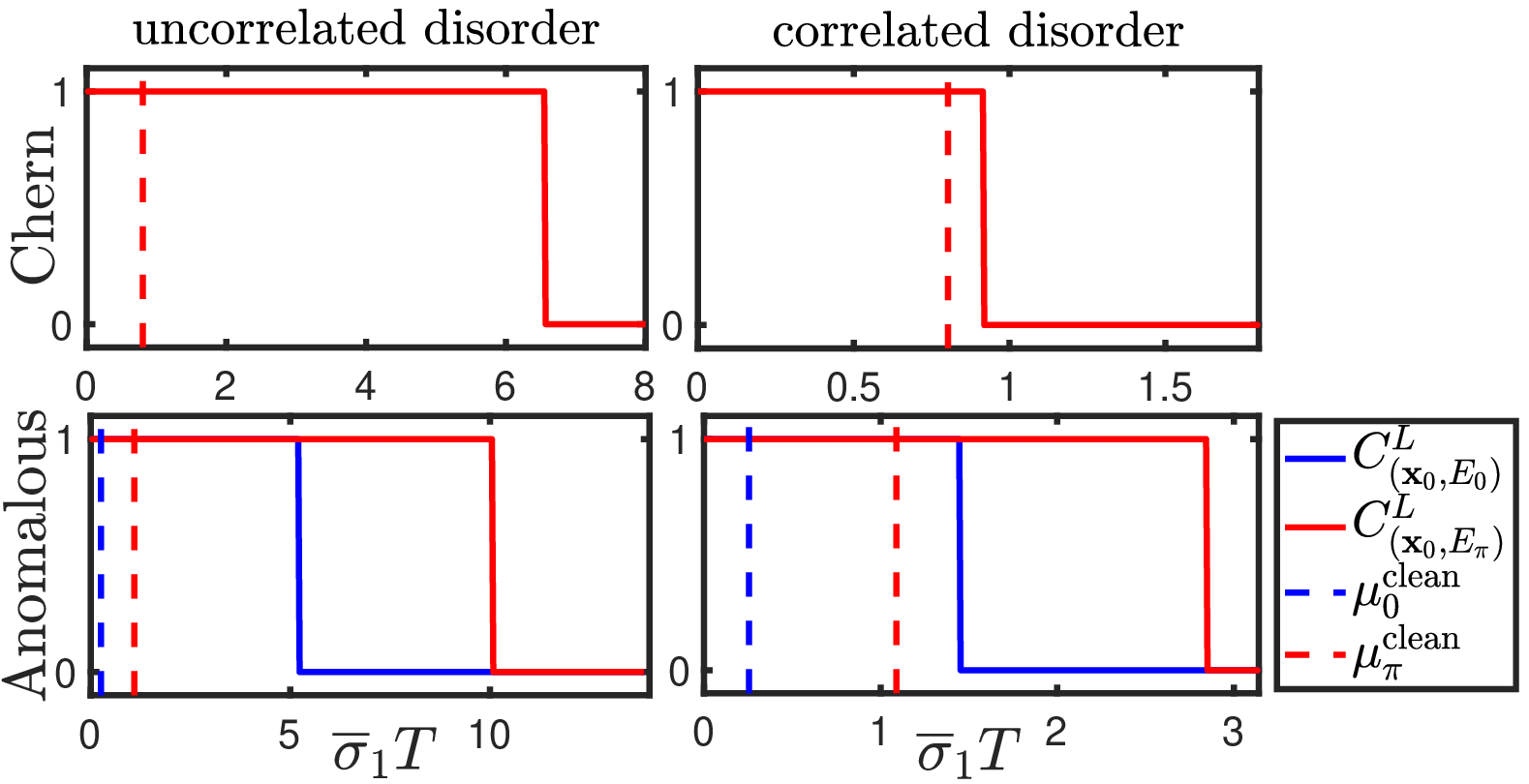}
    \caption{
    Topological  transitions as the coupling strength in the $2^{\rm nd}$ and $4^{\rm th}$ subintervals decreases, and disorder increases. According to (\ref{top_protect_cond}), no phase transition can occur for $\overline{\sigma}_1$ less than $\mu_E^{\rm clean}$ (dashed lines). Transitions induced by uncorrelated disorder (left column) occur at  larger values of $\overline{\sigma}_1$ relative to correlated disorder (right column). The top (bottom) row corresponds  to perturbations of the  Chern (anomalous) insulator shown in Fig.~\ref{drive_protocol}(d) (Fig.~\ref{drive_protocol}(f)). The localizer parameters used are: $E_{0,\pi} = 0,\pi/T$, $\boldsymbol{x}_0$ in center of lattice, and $\kappa = 0.5 / (Ta)$. 
 \label{phase_transition_bound}}
\end{figure}

However, although the bound in Eq.~\eqref{top_protect_cond} is not saturated for the uncorrelated disorder configurations we considered, the derived bound can be approached by spatially correlated disorder configurations.
As an example, consider a perturbation in which the coupling strength $J$ in the second $H_2$ and fourth $H_4$ subintervals are gradually decreased, leaving all other parameters the same as in Fig.~\ref{drive_protocol}.
Similar to increasing the strength of the uncorrelated coupling disorder (e.g., as shown in Fig.~\ref{zero_gap_couple_disorder}), Fig.~\ref{phase_transition_bound} also demonstrates that as the strength of the spatially correlated perturbation increases, eventually a phase transition takes place for both the Chern and anomalous insulators.
Yet, correlated disorders yield topological transitions that are closer to the limit of the our derived bound [Eq.~(\ref{top_protect_cond})] with $\overline{\sigma}_1 / \mu^{\text{clean}}_{0,\pi} \gtrsim 1$, as opposed to the uncorrelated disorders where transitions occur at nearly an order magnitude larger $\overline{\sigma}_1 / \mu^{\text{clean}}_{0,\pi} \sim 10 > 1$. 
The mismatch in the transition thresholds suggests an overestimated bound on the perturbation for uncorrelated disorders and strong topological protection against unbiased random fabrication imperfections. 
Altogether, we have derived a quantitative measure of topological robustness that can be directly related to perturbations of the system during the driving protocol, and as such, larger local gaps in the clean system yield stronger topological robustness, as shown in the Supplementary Material~\ref{topo_protection_pi_gap_sec}~\cite{supp} with the non-trivial Floquet anomalous topology at the $\pi$-gap.


In conclusion, we have developed a general framework, based on the spectral localizer, to classify the topology in Floquet systems directly from the system's Floquet Hamiltonian and quantify their topological protection.
In particular, although the topology is expressed from a stroboscopic point-of-view with the Floquet Hamiltonian, we have derived a relation between topological protection and perturbation of the instantaneous Hamiltonian as in the static case, and show that non-trivial topology survives as long as the time-average of the instantaneous perturbation is smaller than the local gap.
With the derived bound we found the topological protection is more stringent for correlated disorder than uncorrelated disorder, suggesting that designing Floquet systems with waveguides should prioritize well-defined driving steps over minimizing white-noise perturbation in the refractive index or evanescent coupling.
Furthermore, as the spectral localizer for Floquet systems utilized the Floquet Hamiltonian, derived from the monodromy matrix, rather than the Hamiltonian of the governing equations, the spectral localizer formulation is equation-free and provides a data-driven approach to topology where the monodromy matrix can be obtained directly from experiments, by individually exciting each site of the system with unit amplitude at initial conditions to construct the monodromy matrix as in Eq.~(\ref{propagator_exp}), or from some other first principles calculation~\cite{Unal2019}.
Looking forward, we anticipate our operator-based framework should be useful for diagnosing the topology and its topological robustness from a range of physical photonic Floquet systems such as waveguiding or pump-driven systems~\cite{Rechtsman2013, Maczewsky2017, Guglielmon2018, He2019, Ivanov2019, Mukherjee2020, Maczewsky2020a, Maczewsky2020}.

\textit{During the preparation of this manuscript, we became aware of a similar proposal posted on a preprint server~\cite{Ghosh2024}, that also develops a theory of Floquet topological systems based on the spectral localizer.}

\begin{acknowledgments}
S.W.\ acknowledges support from the Laboratory Directed Research and Development program at Sandia National Laboratories.
A.C.\ acknowledges support from the U.S.\ Department of Energy, Office of Basic Energy Sciences, Division of Materials Sciences and Engineering. 
J.T.C.\ acknowledges support from the Air Force Office of Scientific Research under grant FA9550-23-1-0105.
This work was performed, in part, at the Center for Integrated Nanotechnologies, an Office of Science User Facility operated for the U.S. Department of Energy (DOE) Office of Science. Sandia National Laboratories is a multimission laboratory managed and operated by National Technology \& Engineering Solutions of Sandia, LLC, a wholly owned subsidiary of Honeywell International, Inc., for the U.S.\ DOE's National Nuclear Security Administration under contract DE-NA-0003525. The views expressed in the article do not necessarily represent the views of the U.S.\ DOE or the United States Government.
\end{acknowledgments}

\bibliography{Floquet_Insulator_Bib}

\end{document}